\documentclass{svjour3}                     
\smartqed  
\usepackage{graphicx}
\newcommand{\be}{\begin{equation}}
\newcommand{\ee}{\end{equation}}
\newcommand{\bea}{\begin{eqnarray}}
\newcommand{\eea}{\end{eqnarray}}
\journalname{General Relativity and Gravitation}
\begin{document}
\title{Phase transitions in geometrothermodynamics}
\author{Hernando Quevedo\and Alberto S\'anchez\and Safia Taj\and Alejandro V\'azquez}

\institute{
H. Quevedo
\at
Dipartimento di Fisica, Universit\`a di Roma "La Sapienza", Piazzale Aldo Moro 5, I-00185 Roma, Italy;
ICRANet, Piazza della Repubblica 10, I-65122 Pescara, Italy  \\
\email{quevedo@nucleares.unam.mx}  \\
\emph{On sabbatical leave from Instituto de Ciencias Nucleares, Universidad Nacional Aut\'onoma de M\'exico} \\
A. S\'anchez
\at
Instituto de Ciencias Nucleares,
Universidad Nacional Aut\'onoma de M\'exico,
 AP 70543, M\'exico, DF 04510, Mexico \\
\email{asanchez@nucleares.unam.mx} \\
S. Taj
\at
Dipartimento di Fisica, Universit\`a di Roma "La Sapienza", Piazzale Aldo Moro 5, I-00185 Roma, Italy;
ICRANet, Piazza della Repubblica 10, I-65122 Pescara, Italy \\
\emph{Permanent address: Center for Advanced Mathematics and Physics,
National University of Science and Technology, Campus of EME College,
Peshawar Road, Rawalpindi, Pakistan} \\
\email{safiataaj@gmail.com} \\
A. V\'azquez
\at
Facultad de Ciencias,
Universidad Aut\'onoma del Estado de Morelos,
Cuernavaca, MO 62210, Mexico \\
\email{alec\_vf@nucleares.unam.mx}
}

\date{Received: date / Accepted: date}

\maketitle

\begin{abstract}
Using the formalism of geometrothermodynamics, we investigate the geometric properties of the
equilibrium manifold for diverse thermodynamic systems. Starting from Legendre invariant
metrics of the phase manifold, we derive thermodynamic metrics for the equilibrium manifold whose
curvature  becomes singular at those points where phase transitions of first and second order occur.
We conclude that the thermodynamic curvature of the equilibrium manifold, as defined in geometrothermodynamics,
can be used as a measure of thermodynamic interaction in diverse systems with two and three thermodynamic
degrees of freedom.

\end{abstract}

\section{Introduction}
\label{sec:int}

Although thermodynamics is based entirely upon empirical results which are satisfied under certain
conditions in almost any macroscopic system, the geometric approach to thermodynamics
has proved to be useful and illuminating.
In very broad terms, one can say that the following three branches of geometry
have found sound applications
in equilibrium thermodynamics: analytic geometry, Riemannian geometry, and contact geometry.

To illustrate the role of analytic geometry \cite{gibbs,car}, let us consider a thermodynamic system specified by an
equation of state of the form $f(P,V,T)=0$. In the $PVT-$space, the equation of state defines
a surface whose points represent the states in which the system can exist in thermodynamic equilibrium.
Given a surface in a 3-dimensional space, its properties and structure are investigated in the
framework of analytic geometry. Probably, one of the most important contributions of analytic geometry
to the understanding of thermodynamics is the identification of points of phase transitions with
extremal points of the surface (or its projections on the coordinate planes)
determined by the state equation $f(P,V,T)=0$. More detailed descriptions of these contributions can be
found in any good textbook on thermodynamics (see, for instance, \cite{callen,huang}).

Riemannian geometry was first introduced in statistical physics and thermodynamics
by Rao \cite{rao45}, in 1945, by means of a metric whose components in local coordinates coincide with Fisher's
information matrix. Rao's original work has been followed up and extended by a number of
authors (see, e.g., \cite{amari85} for a review). On the other hand, Riemannian geometry in the space of
equilibrium states was introduced by Weinhold \cite{wei75} and Ruppeiner \cite{rup79,rup95},
who defined metric structures as the Hessian of the internal energy and the entropy, respectively.

Contact geometry was introduced by Hermann \cite{her73} into the thermodynamic phase space
in order to take into account the invariance of ordinary thermodynamics with respect
to Legendre transformations, i.e., to the choice of thermodynamic potential.
In order to incorporate Legendre invariance in Riemannian structures
at the level of the phase space and the equilibrium space, the formalism of
geometrothermodynamics (GTD)
\cite{quev07} was recently proposed and applied to different systems. The main motivation for
introducing the formalism of GTD was to formulate a geometric approach which takes into account
the fact that in ordinary thermodynamics the description of a system does not depend
on the choice of thermodynamic potential. One of the main goals of GTD has been to interpret in an
invariant manner the curvature of the equilibrium space as a manifestation of the thermodynamic
interaction. This would imply that an ideal gas and its generalizations with no mechanic interaction
correspond to a Riemannian manifold with vanishing curvature. Moreover, in the case of interacting
systems with non-trivial structure of phase transitions, the curvature should be non-vanishing and
reproduce the behavior near the points where phase transitions occur. These intuitive statements
represent concrete mathematical conditions for the metric structures of the spaces of phase and equilibrium.
The main goal of the present work is to present metric structures which satisfy these conditions for
systems with no thermodynamic interaction as well as for systems characterized by
interaction with first and second order phase transitions.

This paper is organized as follows. In Section \ref{sec:gen} we present the fundamentals of GTD.
In Section \ref{sec:second} we analyze the case of systems with two and three degrees of freedom
which are characterized by second order phase transitions. A metric is derived for the equilibrium
manifold which correctly reproduces the phase transitions in terms of curvature singularities.
Section \ref{sec:first} is dedicated to the investigation of a thermodynamic metric which
reproduces the first order phase transitions of realistic gases and fluids.
 Finally, Section \ref{sec:con} is devoted
to discussions of our results and suggestions for further research.
Throughout this paper we use units in which $G=c=\hbar =k_{_B}=1$.

\section{General framework }
\label{sec:gen}
The starting point of GTD is the
$(2n + 1)$-dimensional thermodynamic phase space ${\cal T}$ which
can be coordinatized by the set of independent coordinates
$\{\Phi,E^{a}, I^{a}\},\ a = 1, ..., n$, where $\Phi$ represents the
thermodynamic potential, and $E^{a}$ and $I^{a}$ are the extensive
and intensive thermodynamic variables, respectively. The positive
integer $n$ indicates the number of thermodynamic degrees of freedom
of the system. Let the fundamental Gibbs 1--form be
defined on ${\cal T}$ as $\Theta = d\Phi - \delta_{ab}I^{a}dE^{b}$, with
$\delta_{ab}={\rm diag}(1,...,1)$. The pair $({\cal T},\Theta)$ is called a
contact manifold \cite{her73,burke} if ${\cal T}$ is differentiable and $\Theta$
satisfies the condition $\Theta\wedge(d\Theta)^{n}\neq0$.
Consider also on ${\cal T}$ a non--degenerate  Riemannian metric $G$
which is invariant with respect to Legendre transformations \cite{arnold}
\begin{equation}
(%
\Phi,E^{a},I^{a})\rightarrow(\tilde{\Phi},\tilde{E^{a}},\tilde{I^{a}}),
\end{equation}
\begin{equation}
 \Phi = \tilde \Phi - \delta_{kl} \tilde E ^k \tilde I ^l \ ,\quad
 E^i = - \tilde I ^ {i}, \ \
E^j = \tilde E ^j,\quad
 I^{i} = \tilde E ^ i , \ \
 I^j = \tilde I ^j \ ,
 \label{leg}
 \end{equation}
where $i\cup j$ is any disjoint decomposition of the set of indices $\{1,...,n\}$,
and $k,l= 1,...,i$. In particular, for $i=\{1,...,n\}$ and $i=\emptyset$ we obtain
the total Legendre transformation and the identity, respectively.
Legendre transformations are a special case of contact transformations which
leave invariant the contact structure of ${\cal T}$. Legendre invariance
guarantees that the geometric properties of $G$ do not depend
on the thermodynamic potential used in its construction.
The set $({\cal T},\Theta,G)$ defines a Riemannian contact manifold which is called
the {\it thermodynamic phase space} (phase manifold), and represents
an auxiliary structure which is necessary to implement in a
consistent manner the properties of ordinary thermodynamics.

The {\it space of thermodynamic equilibrium states} (equilibrium manifold)
is an $n-$dim\-ensional
Riemannian submanifold ${\cal E}\subset {\cal T}$ induced by a smooth map
$\varphi:{\cal E}\rightarrow{\cal  T}$ , i.e. $\varphi:(E^a) \mapsto
(\Phi,E^a,I^a)$ with $\Phi=\Phi(E^a)$ such that
$
\varphi^*(\Theta)=\varphi^*(d\Phi - \delta_{ab}I^{a}dE^{b})=0
$
holds, where $\varphi^*$ is the pullback of $\varphi$. This implies the
relationships
\begin{equation}
 d\Phi
=\delta_{ab}I^{a}dE^{b}\ , \quad \frac{\partial \Phi}{\partial E^{a}}=\delta_{ab}I^{b}\ .
\label{firstlaw}
\end{equation}
The first of these equations corresponds to the first law of
thermodynamics, whereas the second one is usually known as the
condition for thermodynamic equilibrium \cite{callen}. In the GTD formalism,
the last equation also means that the intensive thermodynamic
variables are dual to the extensive ones. Notice that the mapping
$\varphi$ as defined above implies that the equation $\Phi =
\Phi(E^{a})$ must be explicitly given. In standard thermodynamics
this is known as the fundamental equation from which all the
equations of state can be derived \cite{callen,huang}. In this representation,
the second law of thermodynamics is equivalent to the convexity
condition on the thermodynamic potential $\partial ^{2}\Phi /
\partial E^{a}\partial E^{b}\geq 0$, if $\Phi$ coincides with the
internal energy of the system \cite{her73}. The smooth map $\varphi$ induces
in a canonical manner a metric $g$ (thermodynamic metric)
 in the equilibrium manifold ${\cal E}$ by
means of $g=\varphi^*(G)$. Consequently, ${\cal E}$ is a Riemannian manifold
with a non--degenerate metric $g$.

In general, the thermodynamic potential is a homogeneous function of
its arguments, i.e., $\Phi(\lambda E^{a}) = \lambda
^{\beta}\Phi(\lambda E^{a})$ for constant parameters $\lambda$ and
$\beta$. Using the first law of thermodynamics, it can easily be
shown that this homogeneity is equivalent to the relationships
\begin{equation}
\beta\Phi=\delta_{ab}I^{a}E^{b}\ ,\quad
(1-\beta)\delta_{ab}I^{a}dE^{b}+\delta_{ab}E^{a}dI^{b}=0 ,
\label{euler}
\end{equation}
which are known as Euler's identity and Gibbs--Duhem relation,
respectively.

From the above discussion we conclude that to describe a thermodynamic
system in GTD it is necessary to specify a metric $G$ for the phase
manifold ${\cal T}$ and a fundamental equation $\Phi=\Phi(E^a)$. These ingredients
allows us to construct explicitly the pair $({\cal E}, g)$ that constitutes the
equilibrium manifold whose geometric properties should be related to the
thermodynamic properties of the specific system. In the following sections
we will present metrics $G$ which generate equilibrium manifolds for
different thermodynamic systems.


\section{Systems with second order phase transitions}
\label{sec:second}
The non-degenerate metric
\begin{equation}
G = (d\Phi - \delta_{ab}I^{a}dE^{b})^{2} +
(\delta_{ab}E^{a}I^{b})(\eta_{cd}dE^{c}dI^{d}) ,\  \eta_{ab} =
{\rm diag}(-1, 1, ..., 1)
\label{gup1}
\end{equation}
is invariant with respect to total Legendre transformations (\ref{leg}) and,
consequently, can be used to describe the geometric properties of the phase manifold ${\cal T}$.
The smooth map $\varphi: \{E^a\}\mapsto \{ \Phi(E^a),E^a, I^a(E^a)\}$, satisfying $\varphi^*(d\Phi - \delta_{ab}I^{a}dE^{b})=0$,
induces the thermodynamic metric
\be
g=\varphi^*(G)= \left(E^{c}\frac{\partial{\Phi}}{\partial{E^{c}}}\right)
\left(\eta_{ab}\delta^{bc}\frac{\partial^{2}\Phi}{\partial {E^{c}}\partial{E^{d}}} dE^a dE^d \right)\
\label{gdown1}
\ee
that can be explicitly calculated once the fundamental equation $\Phi=\Phi(E^a)$ is specified. Notice that by virtue of the
equilibrium conditions (\ref{firstlaw}) and Euler's identity (\ref{euler}), the conformal factor of the thermodynamic metric
(\ref{gdown1}) turns out to be proportional to the thermodynamic potential $\Phi$. This means that in general we can assume
that the conformal factor is different from zero.

For concreteness we assume that $E^1=S$ is the entropy of a system with two thermodynamic degrees of freedom ($n=2)$.
Then, the dual intensive variable
$I^1 =\partial \Phi/\partial S = T$ represents the temperature, and the thermodynamic metric (\ref{gdown1}) reduces
to
\be g = ( ST + E^2 I^2) \left[-\Phi_{_{SS}} dS^2 + \Phi_{22} (dE^2)^2\right] \ ,
\label{gd2m}
\ee
where a subscript represents derivation with respect to the corresponding coordinate.

To describe the phase transition structure of the system we use the heat capacity $C=T(\partial T/\partial S)=\Phi_{_S}/\Phi_{_{SS}}$,
so that phase transitions of second order occur at those points where $C$ diverges. For the calculation of concrete examples it is
necessary to specify the fundamental equation $\Phi=\Phi(S,E^2)$. In particular, we are interested in analyzing the thermodynamic
properties of black hole configurations \cite{davies} for which usually $\Phi=M$ is the mass (energy) of the black hole and $E^2$ is an additional
extensive variable like electric charge, angular momentum, etc. In this case, the fundamental equation is essentially equivalent to
the entropy--area
relation $S=S(M,E^2)= kA$, where $A$ is the area of the horizon and $k$ is a constant that depends on the dimension of the spacetime.
However, it turns out that it is not always possible to express the fundamental equation in the form $M=M(S,E^2)$ and the
entropy--area relation $S=S(M,E^2)$ must be used. In ordinary thermodynamics this corresponds to a change from the energy representation
to the entropy representation. GTD allows us to perform changes of representations in a simple manner. In fact, to obtain the entropy
representation of the metric (\ref{gd2m}) we rewrite the first law of thermodynamics $d M = T dS + I^2 d E^2$ as $dS =
(1/T) d M - (I^2/T) d E^2$ and identify $M$ and $E^2$ as the new extensive variables, whereas $1/T$ and $-I^2/T$ represent the
corresponding new dual intensive variables. Replacing these new variables in the general expression (\ref{gdown1}), we obtain the
metric
\be
g = \left( {M}S_{_M} + E^2 S_2 \right) \left[ - S_{_{MM}} dM^2 + S_{22} (d E^2)^2\right]\ ,
\label{gd2s}
\ee
which allows us to investigate the properties of the same thermodynamic system in the entropy representation.

\begin{table}[h!]
\begin{center}
\begin{tabular}{|c|c|}
\hline
{\bf Black hole} & {\bf Fundamental equation} \\
\hline
 RN &     $ S=\pi( M+\sqrt{M^2-Q^2} )^2 $  \\
 & \\
 RNAdS &   $ M = \frac{(D-2)\omega_{D-2}}{16\pi}\left(\frac{4S}{\omega_{D-2}}\right)^{\frac{D-1}{D-2}}\left[\frac{1}{l^2}
+ \left(\frac{\omega_{D-2}}{4S}\right)^{\frac{2}{D-2}}+ \frac{2\pi^2Q^2}{(D-2)(D-3)S^2} \right]$         \\
 & \\
Kerr     & $S=2\pi ( M^2+\sqrt{M^4-J^2}) $ \\
 & \\
KAdS     & $ M=\frac{D-2}{2\pi}\left(\frac{\omega_{D-2}}{2^D}\right)^{\frac{1}{D-2}}S^{\frac{D-3}{D-2}}
                      \left(1+\frac{4\pi^2J^2}{S^2}\right)^{\frac{1}{D-2}}$ \\
 & \\
BTZ  &  $ M=\frac{S^2}{16\pi^2 l^2}+\frac{4\pi^2J^2}{S^2}$ \\
 & \\
BTZCS &  $ M=\frac{1}{8\pi^2k^2}\left[ S^2 + 8\pi^2kJ+ \frac{S}{l}\sqrt{(l^2-k^2)(S^2+16\pi^2kJ)}\right]$  \\
 & \\
BTZTF  &  $ S = 2\sqrt{2} \pi l \left[M+\left( M^2-\frac{J^2}{l^2}\right)^{\frac{1}{2}} \right]^{\frac{1}{2}}-
    \frac{3}{2}\ln{ 2\sqrt{2} \pi l \left[ M+\left( M^2-\frac{J^2}{l^2}\right)^{\frac{1}{2}} \right]^{\frac{1}{2}} } $ \\
&  \\
EGB &    $ M=S^{\frac{2}{3}}+\frac{Q^2}{3S^{\frac{2}{3}}} $ \\
 & \\
EGBME   & $ S = \frac{1}{8}\left[ \sqrt{M+\frac{2Q}{\sqrt{3}}} + \sqrt{M-\frac{2Q}{\sqrt{3}}}\right]^3 +
     3\tilde{\alpha}\left[ \sqrt{M+\frac{2Q}{\sqrt{3}}}+ \sqrt{M-\frac{2Q}{\sqrt{3}}}\right] $ \\
 & \\
EYMGB    & $ M=S^{\frac{2}{3}}-\frac{2}{3}Q^2\ln S  $ \\
 & \\
EMGB   & $ M=\pi \alpha+\frac{\pi Q^2}{6S^{\frac{2}{3}}} +\frac{\pi}{2}S^{\frac{2}{3}}-\frac{\pi \Lambda}{12}S^{\frac{4}{3}}$ \\
\hline
\end{tabular}
\caption{{\bf Black holes with two degrees of freedom.} Here we use the following notations: RN = Reissner-Nordstr\"om,
RNAdS =  Reissner-Nordstr\"om-Anti-de-Sitter, KAdS = Kerr-Anti-de-Sitter, BTZ = Ba\~nados-Teitelboim-Zanelli, BTZCS =
Ba\~nados-Teitelboim-Zanelli-Chern-Simons, BTZTF = Ba\~nados-Teitelboim-Zanelli black hole with thermal fluctuations,
EGB = Einstein-Gauss-Bonnet, EGBME = Einstein-Gauss-Bonnet black holes with modified entropy, EYMGB = Einstein-Yang-Mills-Gauss-Bonnet,
EMGB = Einstein-Maxwell-Gauss-Bonnet, $M$ = mass, $S$=entropy, $Q$ = charge, $J$ = angular momentum,
$\Lambda=-1/l^2$= cosmological constant,  and $\omega_{D-2}=2\pi^{\frac{D-1}{2}}/ \Gamma[\frac{D-1}{2}]$ is the volume of the unit
$(D-2)$--sphere,  $k$ = Chern-Simons coupling constant, $\alpha$ = Gauss-Bonnet coupling constant, $\tilde\alpha = (D-2)(D-3)\alpha$.
\protect\label{tab:bhs}
}
\end{center}
\end{table}

The above thermodynamic metrics (\ref{gd2m}) and (\ref{gd2s}) have been used to investigate the black hole configurations
listed in Table \ref{tab:bhs}. In all the cases GTD is mathematically consistent and reproduces the
thermodynamic behavior of the black holes. In fact, the scalar curvature of the equilibrium manifold $({\cal E},g)$
is in all cases different from zero, indicating the presence of non-trivial thermodynamic interaction.
Furthermore, in Table \ref{tab:capcur} we present the heat capacity and the scalar curvature for each
of the black holes contained in Table \ref{tab:bhs}. It follows that at those points where the heat capacity diverges
the scalar curvature of the equilibrium manifold becomes singular. Consequently, a second order phase transition
is characterized by a curvature singularity. This shows that indeed the curvature of the equilibrium manifold
can be considered as a measure of thermodynamic interaction.

The fact that our results are invariant with respect to
Legendre transformations explains some contradictory results \cite{aman06a,caicho99,curflat} that follow
when the equilibrium manifold is equipped with metrics strongly associated to a particular thermodynamic potential.

\begin{table}[h]
\begin{center}
\begin{tabular}{|c|c|c|}
\hline
{\bf BH} & {\bf Heat Capacity } &  {\bf Scalar curvature  } \\
\hline
& & \\
RN & $-\frac{ 2\pi^2r_{+}^2(r_{+ }- r_{-}) }{ (r_{+}-3r_{-}) } $  &
$\frac{(r^2_{+ }-3r_{-} r_{+}+6r_{-}^2)(r_{+}+3r_{-})(r_{+}-r_{-})^2}{\pi^2r_{+}^3(r_{+
}^2+ 3r_{-}^2)^2(r_{+}-3r_{-})^2}$  \\
RNAdS & $\frac{(D-2)S\Big[
\frac{D-1}{l^2}+(D-3)\Big(\frac{\omega_{D-2}}{4S}
\Big)^{\frac{2}{D-2}}-\frac{2\pi^2Q^2}{(D-2)S^2}\Big]}{
\frac{D-1}{l^2}-(D-3)\Big(\frac{\omega_{D-2}}{4S}
\Big)^{\frac{2}{D-2}}+\frac{2(2D-5)\pi^2Q^2}{(D-2)S^2}}$
& $\frac{\mathcal{N}^{RNAdS}} {\left[
\frac{D-1}{l^2}-(D-3)\Big(\frac{\omega_{D-2}}{4S}
\Big)^{\frac{2}{D-2}}+\frac{2(2D-5)\pi^2Q^2}{(D-2)S^2}\right]^2} $\\
& & \\
Kerr &  $\frac{ 2\pi^2r_{+} (r_{+ }+r_{-})^2 (r_{+ }-r_{-}) }{ r_{+}^2-6r_{+}r_{-}-3r_{-}^2} $
     &  $\frac{(3r^3_{+}+3r_{+}^2r_{-}+17r_{+}r_{-}^2+9r_{-}^3)(r_{+}-r_{-})^3}{2\pi^2r_{+}^2
         (r_{+ }+ r_{-})^4( r_{+}^2-6r_{+}r_{-}-3r_{-}^2)^2} $ \\
& & \\
KAdS & $-\frac{(D-2)S[3S^2+20\pi^2J^2-D(S^2+4\pi^2J^2)](S^2+4\pi^2J^2)}{3S^4+24\pi^2J^2S^2+240\pi^4J^4-D(S^4+48\pi^4J^4)}$
     & $\frac{\mathcal{N}^{KAdS}}{[3S^4+24\pi^2J^2S^2+240\pi^4J^4-D(S^4+48\pi^4J^4)]^2}$ \\
& & \\
BTZ & $\frac{ 4\pi r_{+} (r_{+}^2-r_{-}^2)}{ r_{+}^2+3r_{-}^2} $
& $ -\frac{3}{2}\frac{l^4}{(r_+^2+3r_-^2)^2}$ \\
& & \\
BTZCS & $\frac{4\pi(l^2-k^2)r_+^2(r_+^2-r_-^2)}{l[lr_+(r_+^2+3r_-^2)-kr_-(r_-^2+3r_+^2)]}$
        & $\frac{{\cal N}^{{BTZCS}}  }{ l^2[lr_+(r_+^2+3r_-^2)-kr_-(r_-^2+3r_+^2)]^2} $   \\
& & \\
BTZTF & $\frac{ 4\pi r_{+} (r_{+}^2-r_{-}^2)}{r_{+}^2+3r_{-}^2}\,,$
& $\frac{(r_+^2-r_-^2)^2(5r_+^4-6r_+^2r_-^2+9r_-^4)}{4\pi^2r_+^4(r_+^2+3r_-^2)^3}$ \\
& & \\
EGB & $-3S\left(\frac{3S^{\frac{4}{3}}-Q^2}{3S^{\frac{4}{3}}-5Q^2}\right)$
    & $-\frac{729}{2}\frac{(3S^7+10S^{\frac{17}{3}}Q^2-4S^{\frac{13}{3}}Q^4)}
                      {2S^{\frac{5}{3}}(3S^{\frac{4}{3}}-5Q^2)^2(3S^{\frac{4}{3}}+2Q^2)^3}$ \\
& & \\
EGBME &  $ -\frac{\sqrt{M}\left( \sqrt{1+\frac{2\beta}{\sqrt{3}}}-
    \sqrt{1-\frac{2\beta}{\sqrt{3}}}\right)\left( 1+\sqrt{1-\frac{4\beta^2}{3}}+4\delta\right)
    \left(1-\frac{4\beta^2}{3} \right)}{12\left( 1-4\beta^2-8\delta+(1+4\delta)\sqrt{1-\frac{4\beta^2}{3}}\right)}$
 & $\frac{ {\cal N}^{EGBME}} {\left[ 1-4\beta^2-8\delta+(1+4\delta)\right]^2\left( 1-\frac{4\beta^2}{3}\right)} $
 \\
EYMGB & $  S\left( \frac{S^{\frac{2}{3}}-Q^2}{Q^2-\frac{1}{3}S^{\frac{2}{3}}}\right)$
      & $-\frac{\mathcal{N}^{EYMGB} }{ \left(Q^2-\frac{1}{3}S^{\frac{2}{3}} \right)^2\left( Q^2S^{\frac{1}{3}}
      +2Q^2S^{\frac{1}{3}}\ln{ S }-S\right)^3 }$  \\
EMGB  & $ 3S\left( \frac{3S^{\frac{4}{3}}-\Lambda S^2-Q^2}{5Q^2-3S^{\frac{4}{3}}-\Lambda S^2}\right)$
      & $-\frac{  \mathcal{N}^{EMGB} }{\left(2Q^2+3S^{\frac{4}{3}}-\Lambda
S^2\right)^3\left( 5Q^2-3S^{\frac{4}{3}}-\Lambda S^2\right)^2}$ \\
\hline
\end{tabular}
\caption{{\bf GTD of black holes with two degrees of freedom.} Here
we use the following notations: $r_+$ = radius of the exterior
horizon, $r_-$ = radius of the interior horizon, $\beta = Q/M$,
$\delta = \tilde\alpha / M$. The function ${\cal N}$ represents in
each case the numerator of the scalar curvature which is a
well--behaved function at the points where the denominator vanishes.
The results for charged rotating BTZ black holes are generalized in
\cite{asq09} For more details see
Refs.\cite{quev08grg,aqs08,qs08,qs09,qt09}.
\protect\label{tab:capcur} }
\end{center}
\end{table}

The 4--dimensional Kerr-Newman black hole represents a thermodynamic system with three degrees of freedom.
Indeed, from the entropy-area
relation one can derive the fundamental equation  \cite{davies}
\be
S= \pi\left( 2 M^2 - Q^2 +2\sqrt{M^4-M^2Q^2 - J^2}\right) \ ,
\label{entropy}
\ee
which depends on the extensive variables $M$, $J$ and $Q$. The second order
phase transitions are determined by the corresponding heat capacity:
\be
C = -\frac{4TM^3S^3}
{2M^6-3M^4Q^2-6M^2J^2+Q^2J^2 + 2(M^4-M^2Q^2 - J^2)^{3/2} } \ .
\label{capkn}
\ee

According to the general expression for the metric $g$, as given in Eq.(\ref{gdown1}),
the equilibrium manifold is 3--dimensional and the corresponding Legendre invariant
metric reduces to
\be
g^{_{KN}}_{ab}= (MS_{_M} + QS_{_Q}+ J S_{_J})\left(
\begin{array}{ccc}
S_{_{MM}}& 0 & 0 \\
0 & - S_{_{QQ}} & - S_{_{QJ}} \\
0 & - S_{_{QJ}} & - S_{_{JJ}}
\end{array}
\right) \ .
\label{gkn1}
\ee
Inserting here
the expression for the entropy (\ref{entropy}), we obtain a rather cumbersome
metric which cannot be written in a compact form. Nevertheless,
the scalar curvature can be shown to have the form $R^{{KN}} = { {\cal N}^{KN}}/{{\cal D}^{KN}}$, where
\be
{\cal D}^{KN}= \left[2M^6-3M^4Q^2-6M^2J^2+Q^2J^2 + 2(M^4-M^2Q^2 - J^2)^{3/2}\right]^2\ ,
\ee
and ${\cal N}^{KN}$ is a function which is always positive in the black-hole region $M^4\geq M^2Q^2+J^2$.
Since the denominators of the heat capacity and the scalar curvature coincide, we conclude
that there exist curvature singularities at those points where second order phase transitions
occur. This result reinforces the interpretation of the curvature of the manifold ${\cal E}$ as
a measure of thermodynamic interaction.

An additional test for the thermodynamic metrics (\ref{gd2m}) and (\ref{gd2s}) consists in calculating the curvature
of systems with no thermodynamic interaction or no phase transitions. Table \ref{tab:ord} contains the fundamental equations
for some ordinary thermodynamic systems which we investigated from the point of view of GTD.

\begin{table}[h!]
\begin{center}
\begin{tabular}{|c|c|}
\hline
{\bf System} & {\bf Fundamental equation}  \\
\hline
 &  \\
IG  &    $ S=N\left( \ln\frac{V}{N}+\frac{3}{2}\ln\frac{U}{N}\right)$   \\
 &  \\
PaIG &  $ S =  N\left[ \ln{ \frac{V}{N} } + \frac{3}{2} \ln{\frac{U}{N}} -
 \frac{3}{2}\ln{\left(1-\frac{2\mathcal{M}^2}{N^2} \right)} -
\frac{3}{\mu^2} \frac{\mathcal{M}^2}{N^2} \right] $ \\
 &  \\
1IM &     $ F=- J - T \ln\left(\cosh{\frac{H}{T}}+
\sqrt{\sinh^2{ \frac{H}{T} }+e^{-4\frac{J}{T}}} \right) $  \\
\hline
\end{tabular}
\caption{{\bf Ordinary thermodynamic systems.} We use the following notations: IG = ideal gas,
PaIG = paramagnetic ideal gas, 1IM = 1--dimensional Ising model, $U$ = energy, $V$ = volume,
$N$ = total number of molecules, $\mathcal{M}$ = magnetization, $H$ = magnetic field,
$T$ = Temperature,   $J$ = spin interaction parameter, $F$ = Helmholtz free energy.
\protect\label{tab:ord}}
\end{center}
\end{table}

Table  \ref{tab:ord1} contains
the results of our analysis.
In the case of the ideal gas and its paramagnetic generalization, which have no thermodynamic interaction, the curvature
vanishes and the equilibrium manifold becomes flat. In fact, one can show the general result that any generalization of the ideal
gas whose fundamental equation can be separate in its variables as $S=S_1(U)+S_2(V)+S_3(E^3)+...$ always
generates a flat thermodynamic metric \cite{vqs09}.
The 1--dimensional Ising model \cite{statistics}
generates a metric whose curvature is non--zero, indicating the presence of thermodynamic interaction, and
regular everywhere, indicating the lack of second order phase transitions.
Consequently, GTD reproduces at the level of the
curvature the properties associated with the thermodynamic interaction between the particles of the mentioned thermodynamic systems.

\begin{table}[h]
\begin{center}
\begin{tabular}{|c|c|c|}
\hline
{\bf System } & {\bf Heat Capacity } & {\bf Curvature scalar } \\
\hline
& & \\
IG  & $\frac{3}{2}N$ &  $R=0$  \\
& & \\
PaIG
& $\frac{3}{2}N\left( 1-\frac{2\mathcal{M}^2}{N^2}\right)$  & $R=0$\\
& & \\
1IM
& $-\frac{1}{T^2}\left\{\frac{M(T,H)}{I^3\left[I+\cosh \frac{H}{T} \right]}
 -\frac{N(T,H)^2}{I^2\left[I+\cosh{\frac{H}{T}}\right]^2}\right\}$ & $R\neq 0$\\
 & & \\
\hline
\end{tabular}
\caption{{\bf GTD of some ordinary systems.} Here
$I=I(T,H,J)$ is function with non-zero values. The Ricci scalar for the Ising model cannot be put in a compact form, but
a numerical analysis shows that it is everywhere regular.
\protect\label{tab:ord1}
}
\end{center}
\end{table}

\section{Systems with first order phase transitions}
\label{sec:first}
Consider the non--degenerate and Legendre invariant metric
\begin{equation}
G = (d\Phi - \delta_{ab}I^{a}dE^{b})^{2} +
(\delta_{ab}E^{a}I^{b})(\delta_{cd}dE^{c}dI^{d}) ,
\label{gup2}
\end{equation}
for the phase manifold ${\cal T}$.  This metric
is invariant with respect to total Legendre transformations (\ref{leg}) and,
consequently, can be used to describe the geometric properties of ${\cal T}$.
If we assume that the equilibrium manifold ${\cal E} \subset {\cal T}$ has as coordinates the extensive
variables $\{E^a\}$, the smooth embedding map ${\cal E}\rightarrow {\cal T}$,
satisfying $\varphi^*(d\Phi - \delta_{ab}I^{a}dE^{b})=0$, can be used to generate in a canonical way the thermodynamic metric
\be
g=\varphi^*(G)= \left(E^{c}\frac{\partial\Phi}{\partial E^c} \right)
\left(\frac{\partial^{2}\Phi}{\partial {E^{a}}\partial{E^{b}}} dE^a dE^b \right)\
\label{gdown2}
\ee
for the equilibrium manifold ${\cal E}$. In the case of a thermodynamic systems with only two degrees of freedom $(n=2$),
let us introduce the extensive variables $\Phi=U=$ internal energy, $E^1=S=$ entropy, and $E^2=V=$ volume, together with the
intensive variables $T=$ temperature, and $P=$ pressure. The first law of thermodynamics (\ref{firstlaw}) reads $dU = TdS - P dV$
and the explicit form of the thermodynamic metric is
\be
g= (ST - PV )\left( U_{_{SS}} dS^2 + 2 U_{_{SV}} dS dV + U_{_{VV}} d V^2\right)\
\label{gdow22}
\ee
or, equivalently, in the entropy representation
\be
g= \frac{1}{T}(U+VP) \left( S_{_{UU}} dU ^2 + 2 S_{_{UV}} dU dV + S_{_{VV}} dV^2\right)\ .
\ee

Probably, the best--known thermodynamic system with a very rich structure of first order phase transitions is the
van der Waals fluid \cite{callen}. For the sake of generality, we will use here the van der Waals fundamental equation
together with the theorem of corresponding states in order to recast the fundamental equation in an invariant form
applicable to all fluids
\be
\overline{S}=\ln\left( 3\overline{V}-1\right) +\frac{3}{2} \ln\left(\overline{U}+\frac{3}{\overline{V}}\right)\ ,
\label{fevdW}
\ee
where $\overline{U}=U/U_c$ and $\overline{V}=V/V_c$, with $U_c=4a/9b$, $V_c= 3b$, and $a$ and $b$ are the van der Waals constants.
The heat capacity following from this fundamental equation turns out to be constant and, consequently, no second order phase transitions
can occur. However, the critical points determined by the roots of the equation
\be
\overline{P}\,\overline{V}^3-3\overline{V}+2  =0\, ,
\label{crp}
\ee
correspond to first order phase transitions \cite{callen}, where $\overline P = P/P_c$, with $P_c = a/27b^2$, is  the reduced pressure.

From the fundamental equation (\ref{fevdW}) it is then straightforward to compute the thermodynamic metric in the entropy representation:
\bea
\label{vdw}
g^{_{vdW}} =&-&\frac{9\left (5\overline{U}\,\overline{V}^2-\overline{U}\,\overline{V}-3\overline{V}+3
\right)}{4(3\overline{V}-1 )(\overline{U}\,\overline{V}+3)^3
\overline{V}^2}\Bigg\{\overline{V}^4d\overline{U}^2-6\overline{V}^2d\overline{U}d\overline{V}+\nonumber
\\
&+& 9\Bigg[\frac{2(\overline{U}\,\overline{V}+3)(\overline{U}\,\overline{V}^4-6\overline{V}^2+6\overline{V}-1)}{3(3\overline{V}-1)^2}-1
\Bigg] d\overline{V}^2 \Bigg\} \ .
\eea
The 2--dimensional equilibrium manifold turns out to be curved in general, indicating that the particles of the fluid interact thermodynamically.
Furthermore, the scalar curvature  of the above metric can be written in the form
\be
R= \frac{{\cal N}^{vdW}}{\left(\overline P \,\overline V ^3 - 3\overline V + 2\right)^2 }
\ee
where ${\cal N}^{vdW}$  is a function of $\overline U$, and $\overline V$
that is well--behaved at the points where the denominator vanishes. We see that
the scalar curvature  diverges at the critical points determined by Eq.(\ref{crp}). Consequently, a first order phase transition
can be interpreted geometrically as a curvature singularity. This is in accordance with our intuitive interpretation of thermodynamic
curvature.

It is worth noticing that the metric (\ref{gdown2}) can also be used
to describe the properties of ordinary systems. Indeed, a straightforward computation
of the corresponding thermodynamic metrics, using the fundamental equations of the ordinary
systems listed in Table \ref{tab:ord}, leads to results equivalent to those reviewed in Table \ref{tab:ord1}.

\section{Conclusions}
\label{sec:con}

Geometrothermodynamics is a differential geometric formalism whose objective is to describe in an invariant manner
the properties of thermodynamic systems in terms of geometric concepts. To a specific thermodynamic system, GTD assigns
a particular equilibrium manifold ${\cal E}$ whose geometric properties should be related to the thermodynamic properties
of the system. In particular, we expect that the curvature of the equilibrium manifold is related to the thermodynamic
interaction between the particles that constitute the system.

In this work, we analyzed two different thermodynamic
metrics which can be canonically derived from the non-degenerate Legendre invariant metric
\begin{equation}
G = (d\Phi - \delta_{ab}I^{a}dE^{b})^{2} +
(\delta_{ab}E^{a}I^{b})(\chi_{cd}dE^{c}dI^{d})\
\end{equation}
of the phase manifold ${\cal T}$, where $\chi_{ab}$ is an arbitrary constant diagonal tensor.
If $\chi_{ab} = \eta_{ab}$, the resulting thermodynamic metric can be used to describe
the properties of systems characterized by second order phase transitions. This was shown in particular
for several black hole configurations in diverse theories and dimensions. In the Euclidean case, $\chi_{ab}=\delta_{ab}$, the
corresponding thermodynamic metric was shown to correctly reproduce the structure of phase transitions of first order in the
specific case of a fundamental equation which describes realistic gases and fluids.

The formalism of GTD indicates that phase transitions occur at those points where the thermodynamic curvature is singular.
The singularities represent critical points where the geometric description of GTD does not hold anymore and must give place
to a more general approach. In ordinary thermodynamics the situation is similar; near the points of phase transitions equilibrium
thermodynamics is not valid and non-equilibrium models must be implemented. It would be interesting to study non-equilibrium thermodynamic
models in the context of a generalized GTD.

Our results show that the metric structure of the phase manifold ${\cal T}$ determines the type of systems that can be described
by a specific thermodynamic metric: An Euclidean structure describes systems with first order phase transitions, whereas a pseudo-Euclidean
structure describes systems with second order phase transitions. At the moment we do not have an explanation for this result.
It is clear that the phase manifold contains information about thermodynamic systems; however, it will be necessary to further
explore its geometric properties in order to understand where the thermodynamic information is encoded. In this context, the use
of variational principles could be of interest \cite{vqs09}.

\begin{acknowledgements}

H. Quevedo and S. Taj would like to thank ICRANet for support.
\end{acknowledgements}


\end{document}